\documentclass[11pt]{article}
\usepackage{moriond}
\usepackage{textcomp}

\def\Journal#1#2#3#4{{#1} {\bf #2}, #3 (#4)}

\def\AP{{\em Astropart. Phys.}}

\def\be{\begin{equation}}
\def\ee{\end{equation}}
\def\bea{\begin{eqnarray}}
\def\eea{\end{eqnarray}}

\newcommand{\Photo}{}

\begin{document}
\vspace*{4cm}
\title{Search for Dark Matter with Liquid Argon and Pulse Shape Discrimination: Results from DEAP-1 and Status of DEAP-3600}

\author{ P. Gorel, for the DEAP Collaboration }

\address{Centre for Particle Physics, University of Alberta, Edmonton, T6G2E1, CANADA}

\maketitle
\abstracts{
In the last decade, Direct Dark Matter searches has become a very active research program, spawning dozens of projects world wide and leading to contradictory results. It is on this stage that the Dark matter Experiment with liquid Argon and Pulse shape discrimination (DEAP) is about to enter. With a 3600~kg liquid argon target and a 1000~kg fiducial mass, it is designed to run background free during 3 years, reaching an unprecedented sensitivity of 10$^{-46}$~cm$^2$ for a WIMP mass of 100 GeV. In order to achieve this impressive feat, the collaboration followed a two-pronged approach: a careful selection of every material entering the construction of the detector in order to suppress the backgrounds, and optimum use of the pulse shape discrimination (PSD) technique to separate the nuclear recoils from the electronic recoils. Using the experience acquired ffrom the 7~kg-prototype DEAP-1, a 3600~kg detector is being completed at SNOLAB (Sudbury, CANADA) and is expected to start taking data in mid-2014.}

\section{Dark matter detection using liquid argon}
For decades, liquid argon (LAr) has been a candidate of choice for large scintillating detectors. It is easy to purify and boasts a high light yield, which makes it ideal for low background, low threshold experiments. The scintillating process involves the creation of excimer, either directly or through ionization/recombination. The ultra-violet photons are emitted when the excited state decays to the ground state, breaking the molecule. Since the wavelength are not energetic enough to recreate the excimer, argon is transparent to its scintillation light, making it suitable for use in very large detectors.

The two energy states are populated depending on the ionizing particle Linear Energy Transfer. Since they have very different time constants (see Table~\ref{tab:ScintProp}), a Pulse Shape Discrimination (PSD) is possible between ionizations due to electrons ($\beta$ or $\gamma$ radiation), or heavier particles (nuclear recoil or $\alpha$). Since the expected WIMP signal is a nuclear recoil, it can be discriminated from the electromagnetic background, especially the high rate $\beta$-decay of $^{39}$Ar isotope, a naturally occurring isotope of argon (1~Bq/kg in natural argon)\cite{PSD}.

\begin{table}[htb]
\caption{Scintillation properties of argon}
\label{tab:ScintProp}
\vspace{0.4cm}
\begin{center}
\begin{tabular}{|c|c|c|}
\hline
 & Singlet & Triplet \\
\hline
Time constant & $\sim$7~ns & $\sim$1.6~\textmu s \\
\hline
Population ratio for Electron ionizing & $33\%$ & $67\%$  \\
\hline
Population ratio for Nucleus ionizing & $75\%$ & $25\%$  \\
\hline
\end{tabular}
\end{center}
\end{table}

\section{DEAP-1}
\begin{figure}[htb]
\centering
\def\svgwidth{450pt}
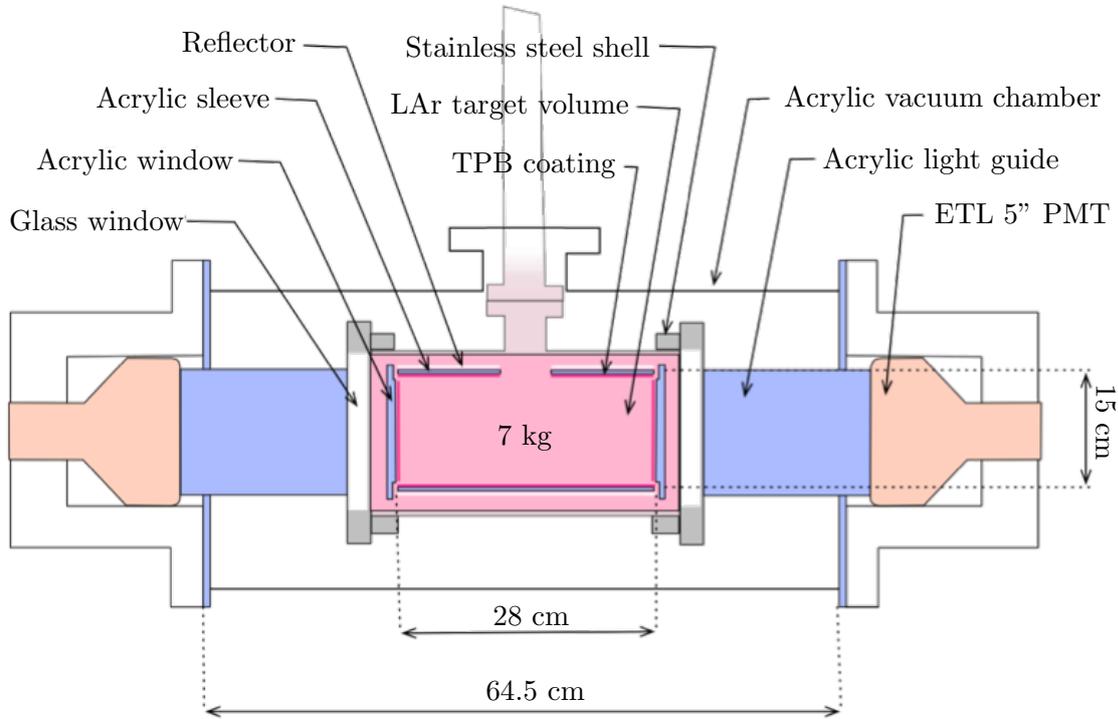
\caption{First version of the DEAP-1 detector: the LAr (7~kg) is contained in a stainless steel shell with an acrylic sleeve. The light guides act both as a thermal insulation and as a shield against neutrons coming from the PMT glass. The TetraPhenylButadiene (TPB) is a wavelength shifter converting the VUV photons emitted by argon into visible light that can travel through the windows and the light guides and be detected by the PMTs.}
\label{fig:DEAP1Det}
\end{figure}

In order to study the background of a LAr dark matter detector and characterize the PSD, the DEAP collaboration built a small scale prototype, DEAP-1, with 7~kg of target material, 2 photomultipliers (PMTs), and a simple geometry (Fig.~\ref{fig:DEAP1Det}). This detector took data between 2007 and 2011, with several iterations mainly concerning with reducing detector backgrounds.

A more detailed review of these results can be found in two papers under publication~\cite{DEAP_Bckgd}~\cite{DEAP_PSD}. The conclusion of this study was twofold: the collaboration acquired a good understanding of the backgrounds in DEAP-1 and confirmed that the key to improve the PSD performance is to maximize the light collection.

The remaining background can be fully explained by a the leakage of the PSD at low energy and by the surface $\alpha$ events at higher energies. The main source of background, coming from the radon decay chains, was reduced to $\sim$16~\textmu Bq/kg for $^{222}$Rn and $\sim$2~\textmu Bq/kg for $^{220}$Rn. Other contributions to the $\alpha$ backgrounds are below 3.5x10$^{-5}$ Hz. 

The data confirmed that the light collection is the main parameter of the PSD performance, and that a light collection of 8 photo-electrons/keV would improve the PSD down to $10^{-10}$ for a 60~keV threshold on the nuclear recoil. These results would suggest that a tonne-scale detector would have less than one leakage in the WIMP region of interest for 3 years of data.

\section{DEAP-3600: Status}
\begin{figure}[htb]
  \centering
  \def\svgwidth{385pt}
  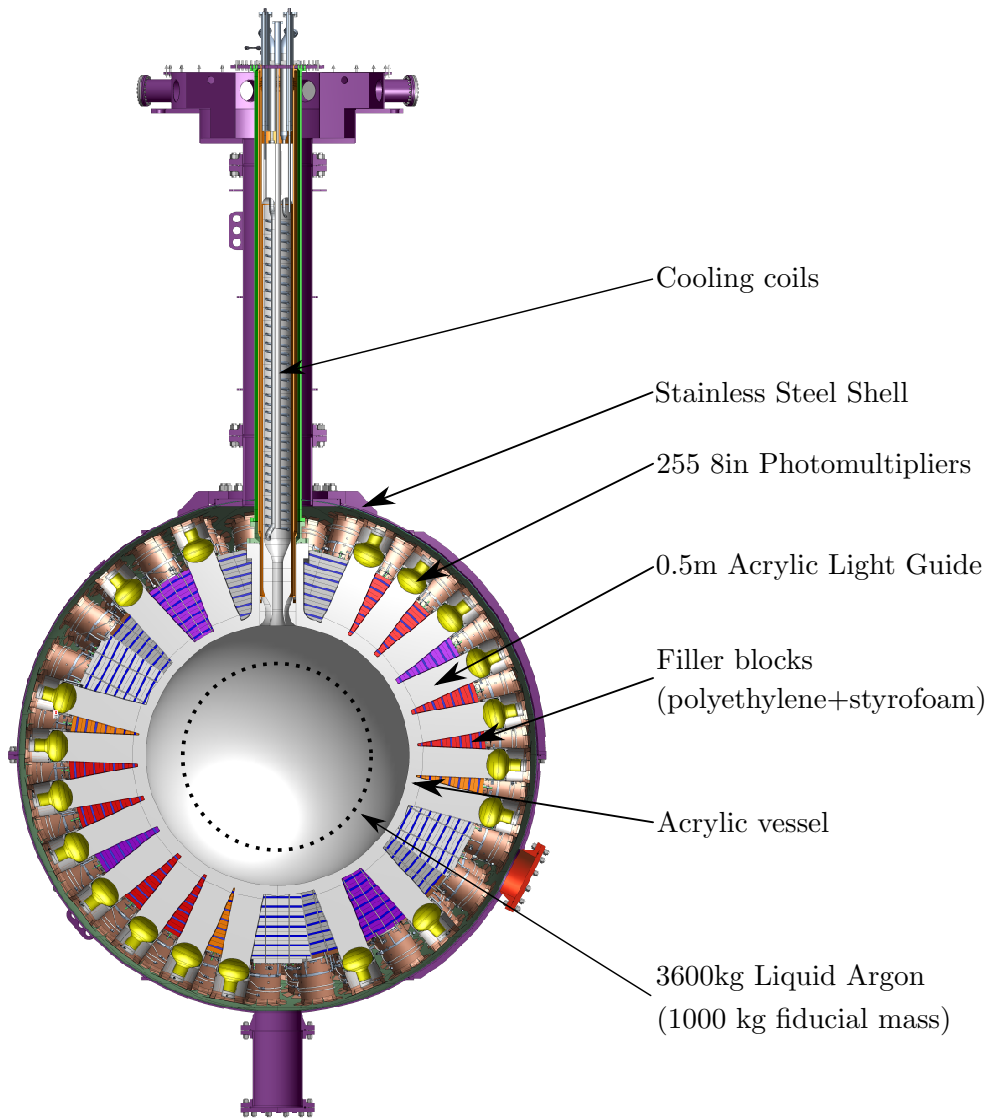
  \caption{Section of the DEAP-3600 detector. The acrylic vessel has in inner radius of 85~cm and holds 3600~kg of LAr, which is viewed by 255 8~inch high quantum efficiency photomultiplier tubes (PMTs) through 50~cm lightguides. The lightguides provide neutron shielding and thermal insulation between the cryogenic acrylic vessel and �warm� PMTs. The inner detector is housed in a large stainless steel spherical vessel, which itself is immersed in an 7~m diameter water shielding tank (not showed).}
  \label{fig:DEAP3600Det}
\end{figure}

DEAP-3600 is a tonne-scale LAr dark matter detector with a total mass of 3,600~kg of target material, and a fiducial mass of 1,000~kg (Fig.~\ref{fig:DEAP3600Det}). The goal is to reach a sensitivity for the WIMP elastic scattering cross section of 10$^{-46}$~cm$^2$ for a  WIMP mass of 100~GeV~\cite{TAUP2011}. 
The detector specifications directly follow the guidelines defined by the DEAP-1 results: firstly the mitigation of background and secondly the maximization of the light collection. 

The first requirement is fulfilled by several layers of shielding coupled with a careful choice of all the material used for the construction. The detector is being built at SNOLAB~\cite{SNOLAB} (Canada) where the rock overburden is equivalent to 6010~m of water and the cosmic muon flux was measured to be 0.23~muon/m$^2$/day. It is located in a 7~m diameter by  7~m height ultra-pure water tank to shield against the rock radioactivity. Between the target material and the light detector, 0.5~m of acrylic light guides and filler blocks (layers of polyethylene and styrofoam) act both as thermal insulation and as a shield against neutrons emitted by $(\alpha,n)$ reactions in the PMT glass. The acrylic vessel, containing the LAr, has been cast out of pure, distilled monomer ensuring a high radiopurity. It should be noted that prior to the evaporation of the purified TPB onto the inner surface, a 1~mm thick layer of acrylic will be sanded out by a custom-made robotic arm in order to remove the radon daughters embedded on and under the surface during the construction.

The second requirement is achieved through the choice of a single phase design which maximizes the photocathode coverage ($\sim75\%$) with 255 8" PMTs. The efficiency is improved by the choice of high quantum efficiency Hamamatsu R5912. In order to gather as much light as possible, the light guides have been glued to the acrylic vessel and wrapped with specular reflectors. 

The construction of the detector is almost finished. The AV is completed, most of the PMTs are mounted and the collaboration is getting ready to place everything inside the steel shell. The electronics and cooling systems are being tested. Commissioning is expected to happen this summer with first data scheduled this fall.

\end{document}